\documentclass[a4paper,11pt]{article}
\pdfoutput=1 

\usepackage{jheppub} 

\usepackage[T1]{fontenc} 

\title{\boldmath Dynamics of photons and shadows for black holes haired with parity-odd fields}

\author[a]{Yang Huang}
\author[b]{Dao-Jun Liu}
\author[a]{Hongsheng Zhang}

\affiliation[a]{School of Physics and Technology, University of Jinan, Jinan 250022, China}
\affiliation[b]{Department of Physics, Shanghai Normal University, Shanghai 200234, China}

\emailAdd{sps\_huangy@ujn.edu.cn}
\emailAdd{djliu@shnu.edu.cn}
\emailAdd{sps\_zhanghs@ujn.edu.cn}

\abstract{Strong self-gravitational fields enable the realization of macroscopic odd-parity quantum objects. Using ray-tracing methods, we systematically analyze the dynamics of photons and the shadow features of rotating black holes with parity-odd scalar hair and contrast them with those of Kerr black holes. Our results demonstrate measurable distinctions between scalar-haired black hole shadows and their Kerr counterparts. Notably, even for tiny scalar charge and negligible scalar hair mass contributions, these differences remain quantitatively resolvable. In particular, one of the hairy black hole reported here lies within the Event Horizon Telescope observational uncertainties, probing the scalar masses of $1.02\times10^{-20}$eV with M87*. These findings may provide related theoretical benchmarks for future observational campaigns targeting scalar-field dark matter candidates through black hole shadow imaging. }

\begin{document} 
\maketitle
\flushbottom

\section{Introduction}
\label{sec:intro}

Black holes (BHs) play a key role in many astrophysical processes, such as the extreme bursting of gravitational wave in binary BH coalescence and the formation of supermassive BHs in the early universe. The study of BHs offers an exciting opportunity to test theories of gravity under extreme conditions. 

According to the uniqueness theorem, stationary BHs in asymptotically flat space-times are characterized by only three observable parameters: mass, spin and charge (if they exist); see Ref.\cite{Chrusciel2012} for a review. However, the theorem is only valid for the Einstein-Maxwell theory. When more general types of matter fields are coupled to gravity, the theory does not prevent the formation of hairy black holes. It is common that a black hole is surrounded by some matters, such as halo of dark matters and accretion disk of baryon matters. Usually, we describe these matters surrounding the hole by using perfect/viscous fluid approximation. Parity is a fundamental concept in modern physics and one of the basic symmetries of quantum mechanics. Generally speaking, macroscopic objects such as tables, chairs, apples,  the Sun, etc., do not possess strictly defined odd or even parity. Only microscopic objects in pure quantum states have well-defined parity. Macroscopic celestial bodies are usually described by fluid models, but such descriptions break down in strong gravitational fields or near horizons, where a smooth description in terms of density and pressure is no longer appropriate. Therefore, celestial bodies—especially those with strong gravitational fields—should in principle be described by quantum mechanics. The matter field of a hairy black hole, i.e., its “hair,” is described by a wave function and is a pure quantum state, which should therefore possess parity properties. In general relativity, the hair of a black hole can be stabilized via the quantum superradiance of a scalar field, and such configurations are likely to exist in real astrophysical environments. The scalar field is in a pure quantum state, and odd-parity states naturally emerge. This would represent the first strict realization of parity symmetry in a macroscopic object and offers a unique opportunity for humanity to directly observe macroscopic parity symmetry.

When the hair is described by a fundamental field, one of interesting hairy black holes (HBHs) emerges in Einstein-Klein-Gordon theory with a massive complex scalar field, the so-called Kerr BHs with scalar hair \cite{Herdeiro:2014goa,Herdeiro2014,Herdeiro:2015gia}. The key ingredient in the construction of these solutions is the synchronization condition between the angular velocity of the event horizon and the phase frequency of the scalar field, which makes the scalar flux does not get through the event horizon. In the limit of vanishing event horizon, these HBHs reduce to the rotating boson star solutions \cite{Schunck1996,Yoshida1997}. By analogy with rotating boson stars (BSs), there are two kinds of such solutions possessing different parity, namely, Kerr BHs with parity-even and parity-odd scalar hair \cite{Wang2019,Kunz2019}. Compared with the Kerr BHs with  scalar hair possessing even parity\cite{Herdeiro:2014goa}, parity-odd HBHs possess distinctive new features which are related to the shapes of their ergoregions \cite{Kunz2019}. A recent perspective suggests that parity-odd boson stars can be interpreted as equilibrium states of binary boson stars. Based on this understanding, stationary solutions for binary BHs have been identified, where the surrounding parity-odd scalar hair provides a force that balances against gravity \cite{Herdeiro:2023roz}.  In various modified gravity theories, HBHs also exist and may deviate from the standard Kerr/Kerr-Newman BH solutions (see, e.g., Refs.\cite{Herdeiro2015,Volkov2017} for a review).  

The Event Horizon Telescope (EHT), through its imaging of the supermassive BHs, provides unprecedented observational data on BH shadows. These observations offer a thrilling opportunity to test the predictions of general relativity and explore the possibility of HBHs \cite{Vagnozzi:2022moj}. By comparing the observed shadow with theoretical models, constraints can be placed on the parameters of HBHs. Therefore, it is interesting to ask how the presence of the parity-odd scalar hair will affect the shadow of a Kerr BH. 

‌The imprints of scalar hair on BH shadows have been extensivel studied \cite{Zhang:2025xnl,Wang:2025dfn,Cunha:2015yba,Cunha:2016bpi,Cunha:2017wao,Gyulchev:2024iel,Cunha:2016bjh}. The images of BHs and BSs surrounded by matter and their light rings have also been extensively studied \cite{Shen:2023nij,Zhang:2024lsf,Wang:2024uda,Zhao:2025yhy,Zeng:2025nmu,Yang:2025usj,Huang:2024bbs,Cui:2024wvz,Huang:2025jfa,Xu:2024gjs,Hu:2023pyd,Hu:2024cbn,Feng:2024iqj,Wang:2023fge}. In a previous study, we investigated the lensing effects of parity-odd BSs, revealing distinct gravitational distortions and chaotic behavior in their images compared to their parity-even counterparts \cite{Huang:2024gtu}. This paper aims to explore the implications of parity-odd HBHs on observed shadows, examining how the presence of parity-odd scalar hair alters shadow characteristics in contrast to both parity-even HBHs and conventional Kerr BHs. Through a detailed analysis of theoretical models, we seek to uncover the unique features of parity-odd HBHs and their observational signatures. The shadow of a general rotating black hole is discussed in \cite{Meng:2023uws}. Throughout the paper, we use the geometric units $G=c=1$ unless otherwise stated.

This paper is organized as follows. In the next section we present the set-up of black hole with parity-odd hair, introducing the mass and angular momentum for the black hole and the hair respectively. In section III,  we address our numerical methodology to study the dynamics of photons in background of odd-parity-hair black hole and the image of the hairy hole. In section IV, we demonstrate the shadows of black hole haired with parity-odd fields with different parameters and from different orientations. In section V, we conclude this paper.

\section{ Set-up}
Consider the model in which a massive scalar field with mass $\mu$ is minimally coupled to gravity 
\begin{equation}\label{Eq: action}
	S=\int d^4x\sqrt{-g}\left(\frac{R}{16\pi}-\mathcal{L}_m\right),
\end{equation}
where $R$ is the Ricci scalar, and $\mathcal{L}_m=g^{\alpha\beta}\Psi^*_{,\alpha}\Psi_{,\beta}+\mu^2\Psi^*\Psi$ is the scalar Lagrangian, which corresponds to a stress-energy $T_{\alpha\beta}=2\Psi^*_{(,\alpha}\Psi_{,\beta)}-g_{\alpha\beta}\mathcal{L}_m$. Variation of the action (\ref{Eq: action}) with respect to the metric leads to the Einstein equations
\begin{equation}\label{Eq: EOM}
	R_{\alpha\beta}-\frac{1}{2}g_{\alpha\beta}R=8\pi T_{\alpha\beta}.
\end{equation}
The equations of motion for the scalar field is the Klein-Gordon equation $\left(\Box-\mu^2\right)\Psi=0$. It can be shown that the system is invariant under the global $U(1)$ transformation of the scalar field, $\Psi\rightarrow\Psi e^{i\chi}$, where $\chi$ is a constant. As a result, the scalar $4$-current, $j^\alpha=-i\left(\Psi^*\partial^\alpha\Psi-\Psi\partial^\alpha\Psi^*\right)$, is conserved, and the corresponding conserved charge is given by
\begin{equation}
	Q=\int d^3x\sqrt{-g}j^t.
\end{equation}

To obtain spinning axially-symmetric solutions, following \cite{Kunz2019}, we write the metric as
\begin{equation}\label{Eq: ansatzMetric}
	\begin{aligned}
		ds^2=-N_1e^{f_0}dt^2+N_2e^{f_1}\left(dr^2+r^2d\theta^2\right)+N_2e^{f_2}r^2\sin^2\theta\left(d\varphi-Wdt\right)^2,
	\end{aligned}
\end{equation}
where 
\begin{equation}
	N_1=\left(\frac{r-r_h}{r+r_h}\right)^2,\;\;N_2=\left(1+\frac{r_h}{r}\right)^4,
\end{equation}
in which $\{f_0, f_1, f_2, W\}$ are functions of $(r,\theta)$, and $r_h$ labels the horizon radius of the BH. The axially symmetric ansatz for the stationary scalar field is
\begin{equation}\label{Eq: ansatzScalar}
	\Psi=\phi(r,\theta)e^{i(m\varphi-wt)},
\end{equation}
where $w$ is the scalar field frequency and $m=\pm1,\pm2,\cdots$ is the azimuthal harmonic index. Without loss of generality, one can assume $w>0$. In this paper, we focus on solutions with $m=1$ for simplicity.

Substituting the ansatz (\ref{Eq: ansatzMetric}) and (\ref{Eq: ansatzScalar}) into equations of motion yields a set of partial differential equations for $\phi$, $f_i\ (i=0,1,2)$, and $W$. These equations will be solved numerically with appropriate boundary conditions. First, a power series expansion near $r=r_h$ yields the conditions of regularity
\begin{equation}
	\partial_r f_0|_{r=r_h}=\partial_r f_1|_{r=r_h}=\partial_r f_2|_{r=r_h}=\partial_r \phi|_{r=r_h}=0.
\end{equation}
The stationary scalar hair around the horizon is supported by the synchronization condition
\begin{equation}
	w=m\Omega_h,
\end{equation}
where $\Omega_h=W|_{r=r_h}$ is the horizon angular velocity. For asymptotic flat solutions at spatial infinity, the boundary conditions are given by
\begin{equation}
	\begin{aligned}
		f_0|_{r\rightarrow\infty}&=f_1|_{r\rightarrow\infty}=f_2|_{r\rightarrow\infty}=W|_{r\rightarrow\infty}=0,\\
		\phi|_{r\rightarrow\infty}&=0.
	\end{aligned}
\end{equation}

Boundary conditions on the symmetry axis at $\theta=0,\pi$ are determined by the axial symmetry and regularity
\begin{equation}\label{Eq: BC1}
	\begin{aligned}
		\partial_{\theta}f_0|_{\theta=0,\pi}&=\partial_{\theta}f_1|_{\theta=0,\pi}=\partial_{\theta}f_2|_{\theta=0,\pi}=0,\\
		\partial_{\theta}W|_{\theta=0,\pi}&=0,\\
		\phi|_{\theta=0,\pi}&=0.
	\end{aligned}
\end{equation}
Also, the condition of absence of conical singularity yields $f_1 |_{\theta=0}=f_2 |_{\theta=0}$.
It can be shown that all metric functions are symmetric with respect to a reflection on the equatorial plane
\begin{equation}\label{Eq: BC2}
	\begin{aligned}
		\partial_{\theta}f_0|_{\theta=\pi/2}&=\partial_{\theta}f_1|_{\theta=\pi/2}=\partial_{\theta}f_2|_{\theta=\pi/2}=0,\\
		\partial_{\theta}W|_{\theta=\pi/2}&=0.
	\end{aligned}
\end{equation}
Meanwhile, the scalar field can exhibit two possible symmetries, determined by their parity. Specifically, solutions with even parity are symmetric, satisfying $\partial_\theta\phi|_{\theta=\pi/2}=0$, while solutions with odd parity are antisymmetric, satisfying $\phi|_{\theta=\pi/2}=0$.

The total ADM angular momentum $J$ and mass $M$ can be read from the asymptotic sub-leading behaviour of the metric functions
\begin{equation}
	g_{t\varphi}=-\frac{2J}{r}\sin^2\theta+\cdots,\;\;g_{tt}=-1+\frac{2M}{r}+\cdots.
\end{equation}
For a HBH, these quantities can be decomposed into contributions from the event horizon and the scalar hair: $J=J_h+mQ$, and $M=M_h+M_{\Psi}$, where
\begin{equation}
	M_{\Psi}=-\int_\Sigma d^3x\sqrt{-g}(2T^t_t-T).
\end{equation}
Also, the scalar hair can be parametrized by $q\equiv mQ/J\in[0,1]$; $q=0$ corresponds the Kerr limit and $q=1$ is the boson star limit \cite{Herdeiro:2014goa}.

\section{ Numerical methodology} 
In order to study the shadows of parity-odd scalar hair, we first reconstruct these solutions presented in \cite{Kunz2019,Wang2019} through numerical methods akin to that employed in \cite{Huang:2024gtu}. We introduce a new compactified radial coordinate
\begin{equation}
	x=\frac{r-r_h}{r+L},
\end{equation}
which maps the semi-infinite region $[0,\infty)$ to the finite interval $[0,1]$, where $L$ is a positive constant. According to the boundary conditions (\ref{Eq: BC1}) and (\ref{Eq: BC2}), the metric functions $f_i, W$ $(i=0,1,2)$ can be written as \cite{Fernandes:2022gde}
\begin{equation}\label{Eq: Expantion Met}
	u(x,\theta)=\frac{1}{2}u_0(x)+\sum^{N_\theta-1}_{n=1}u_n(x)\cos\left(2n\theta\right),
\end{equation}
and the scalar field be written as 
\begin{equation}
	\phi(x,\theta)=\sum^{N_\theta-1}_{n=0}\phi_n(x)\sin(2n\theta).
\end{equation}
Then, the field equations is discretized on a uniform grid in $x$ with fourth-order finite difference scheme. The resulting system of nonlinear algebraic equations is solved by using the Newton-Raphson method.

\begin{figure}
	\centering	
	\includegraphics[trim=10 160 20 160, clip, width=0.7\textwidth]{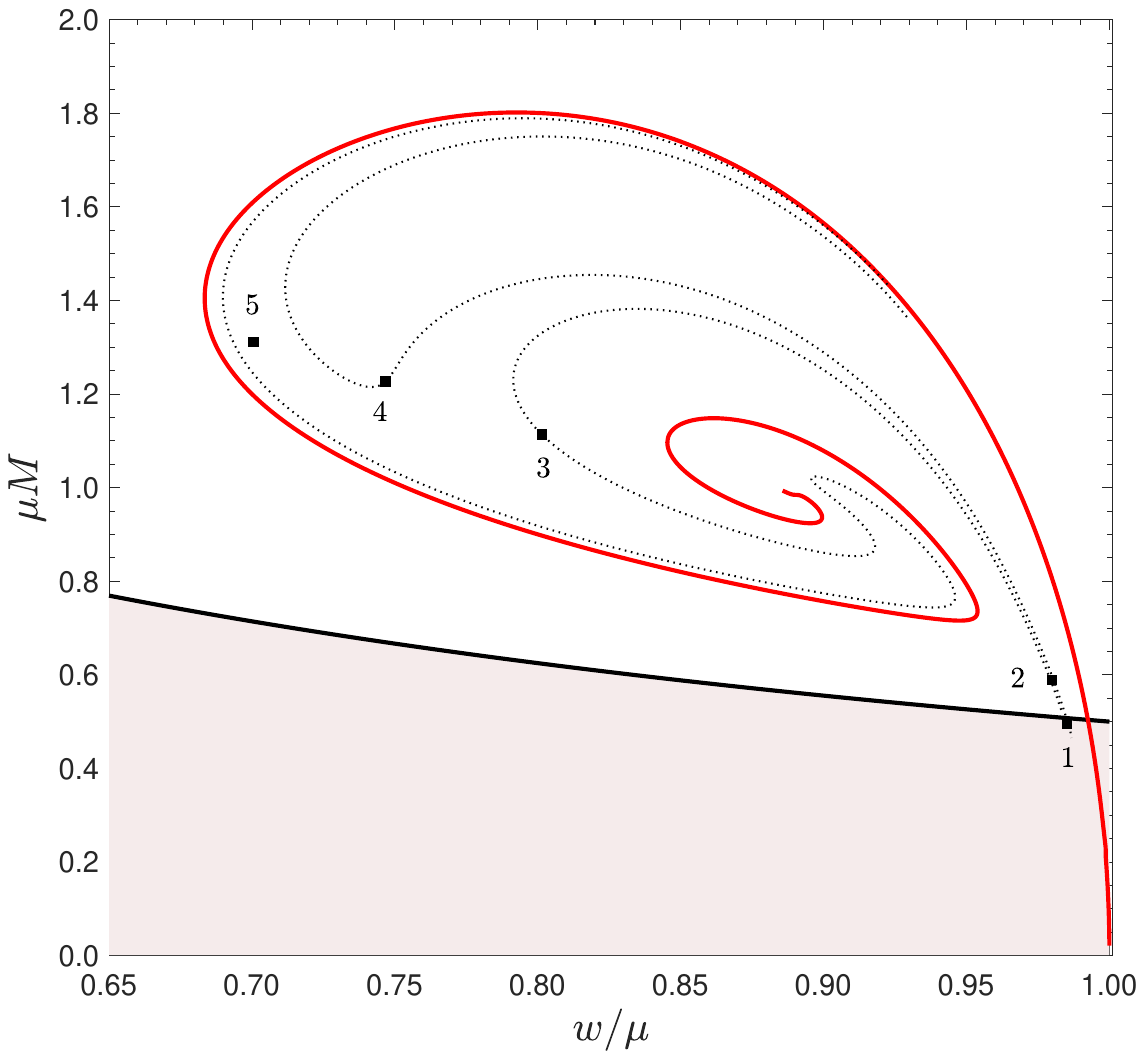}
	\caption{ADM mass $M$ vs frequency $w$ diagram for HBH solutions for $m=1$. Solid black curve: extremal Kerr BHs; solid red curve: parity-odd boson states. Five marked points correspond to 
		solutions studied in detail.}
	\label{Fig: param space}
\end{figure}

The parameter space for the Kerr BHs with parity-odd scalar hair is illustrated in Fig.\ref{Fig: param space}, wherein five distinct solutions have been highlighted for subsequent analysis of BH shadows in the following section. Their physical quantities are presented in Table \ref{Table: Ref1}.

We employ the ray tracing method to study the image of the HBHs. The geodesic motion of a photon is described by the Hamiltonian
\begin{equation}
	\mathcal{H}=\frac{1}{2}g^{\mu\nu}p_\mu p_\nu=0,
\end{equation}
where $p_\mu$ are the $4$-momentum components of the photon. The geodesic equations are given by
\begin{equation}\label{Eq: geodesic}
	\dot{x}^\mu=\frac{\partial\mathcal{H}}{\partial p_\mu},\;\;\;\dot{p}_\mu=-\frac{\partial\mathcal{H}}{\partial x^\mu}.
\end{equation}
For axially-symmetric solutions, there are two Killing vectors, $\partial_t$ and $\partial_\varphi$, which lead to two conserved quantities
\begin{equation}
	E=-p_t,\;\;\;\Phi=p_{\varphi}.
\end{equation}
We adopt the zero angular momentum observer frame. The initial conditions for $\left\{E,p_r,p_\theta,\Phi\right\}$ are determined by the observation angle $(\alpha,\beta)$ as follows 
\cite{Cunha:2016bpi},
\begin{equation}
	\begin{aligned}
		E        &= |\vec{P}|\left(\frac{1+\gamma\sqrt{g_{\varphi\varphi}}\sin\beta\cos\alpha}{\zeta}\right),\\
		p_r      &= |\vec{P}|\sqrt{g_{rr}}\cos\beta\cos\alpha,\\
		p_\theta &= |\vec{P}|\sqrt{g_{\theta\theta}}\sin\alpha,\\
		\Phi     &= |\vec{P}|\sqrt{g_{\varphi\varphi}}\sin\beta\cos\alpha,
	\end{aligned}
\end{equation}
where
\begin{equation}
	\zeta=\sqrt{\frac{g_{\varphi\varphi}}{g^2_{t\varphi}-g_{tt}g_{\varphi\varphi}}},\;\;\gamma=-\frac{g_{t\varphi}}{g_{\varphi\varphi}}\zeta.
\end{equation}
Here, $|\vec{P}|$ only determines the photon's frequency and does not influence the trajectory. 

In the ray tracing method, the distorted image caused by the central object is generated by numerically solving the null geodesic equations for each pixel, with initial conditions specified by $(\alpha, \beta)$. Each pixel is then assigned an appropriate color based on the final state of the corresponding photon. 

\begin{figure}
	\centering
	\includegraphics[trim=0 180 0 180 0, clip, width=0.7\textwidth]{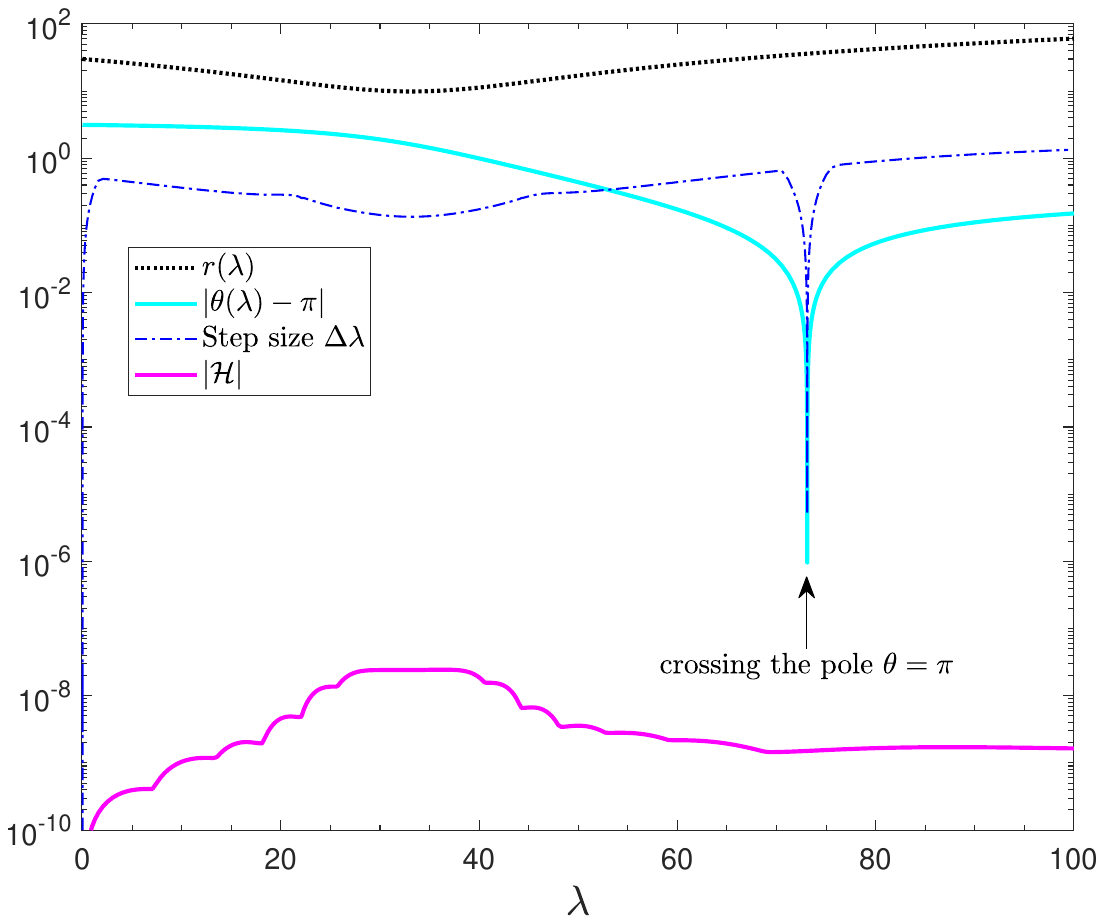}
	\caption{An example of a light ray that crosses the pole at $\theta=\pi$. As the light ray gets closer to the pole, the step size $\Delta\lambda$ is automatically reduced to ensure numerical 
		accuracy. Throughout the crossing, the Hamiltonian constraint is kept below $10^{-8}$.}
	\label{Fig: cross the pole}
\end{figure}

For numerical integration, we employ the Dormand-Prince pair of formulas \cite{DORMAND198019}, in which the step size is dynamically  adjusted based on local error estimates. This approach allows for efficient integration of the vast majority of light rays. However, it encounters challenges when dealing with light rays that pass through the poles at $\theta=0,\pi$, because of the $\left(\sin\theta\right)^{-2}$ terms on the right-hand sides of Eqs.~(\ref{Eq: geodesic}). As the light rays approach these poles, the integration step size becomes excessively small, rendering further integration infeasible. To address this issue, when the light ray is sufficiently close to the poles (e.g. $|\sin\theta|<10^{-6}$), we implement a transformation as follows
\begin{equation}\label{Eq: trans pole}
	\theta, \varphi, p_\theta\rightarrow\left\{
	\begin{aligned}
		2\pi-&\theta, &\varphi+\pi,&-p_\theta,\;\;\text{for}\ \theta>\pi,\\
		-&\theta,     &\varphi+\pi,&-p_\theta,\;\;\text{for}\ \theta<0,\\
		&\theta,      &\varphi+\pi,&-p_\theta,\;\;\text{for}\ 0<\theta<\pi.
	\end{aligned}
	\right.
\end{equation} 
After each integration is completed, we wrap $\varphi$ into the range $[0,2\pi)$. In Fig. \ref{Fig: cross the pole}, we show an example of a light ray crossing the pole at $\theta=\pi$. As the light ray approaches the pole, the step size is automatically decreased to guarantee numerical accuracy. When it gets close enough, the transformation (\ref{Eq: trans pole}) is applied to enable the light ray to cross the pole. Throughout the entire process of the light ray crossing the pole, the Hamiltonian constraint stays at a value below $10^{-8}$, which is the precision limit achievable by interpolation of the numerical solution.

Our ray tracing code is implemented in CUDA C, enabling efficient computation of images for numerical HBH solutions. Typically, an image with a resolution of $2000\times 2000$ pixels can be computed within approximately $2$ hours. Moreover, we find that the majority of the computational time is spent on interpolating the numerical solutions. To verify this point, we test the same code on computing the shadow of an analytical solution (the Kerr BH), where interpolation is unnecessary since the analytical solution can be directly computed. In this case, generating a Kerr image at the same resolution takes only about $2$ minutes.

\begin{figure}
	\centering
	\includegraphics[width=0.45\textwidth]{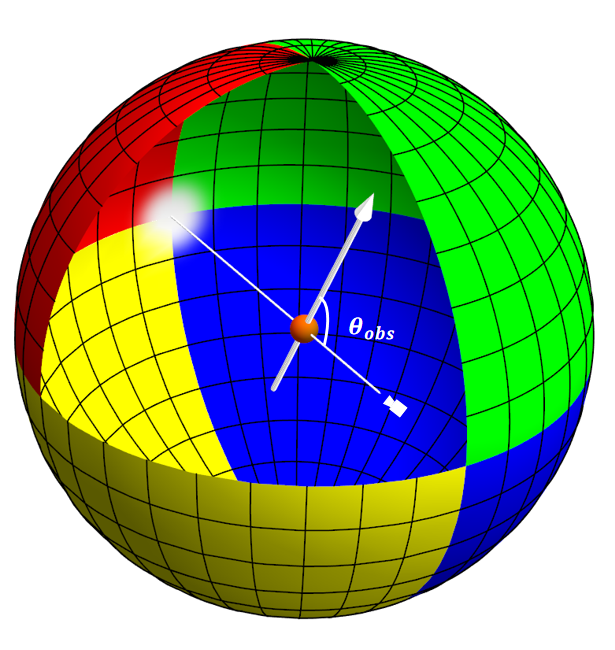}
	\caption{The full celestial sphere with event horizon marked by orange sphere at center.}
	\label{Fig: sphere}
\end{figure}

We employ a setup similar as in the literatures \cite{Huang2018,Chen:2022scf,Cunha:2017wao,Zhang:2025xnl,Liu:2024lve,Hu:2020usx,Sun:2023syd,Chen:2023wzv,Long:2020wqj,Zhang:2022osx,Liu:2024lbi,Liu:2024soc,Yuan:2024wdl,He:2022opa,Guo:2022muy,Zhong:2021mty,Wang:2018eui,Wang:2017qhh,Yang:2024ulu}, as illustrated in Fig. \ref{Fig: sphere}. Specifically, the celestial sphere is divided into four quadrants, each assigned a distinct color based on the angular coordinates of the sphere. Assuming the observer is oriented toward the center of the full celestial sphere, where the BH is located, a white spot on the sphere denotes the observer’s line of sight, while a white arrow indicates the axis of rotation of the BH, which lies in the plane of $0^\circ$ and $180^\circ$ celestial longitude. The angle between the white arrow and the observer’s line of sight, denoted as $\theta_{\text{obs}}$, serves as the initial value of $\theta$ for photons in the ray tracing algorithm. 

\begin{figure}
	\centering	
	\includegraphics[width=0.46\textwidth]{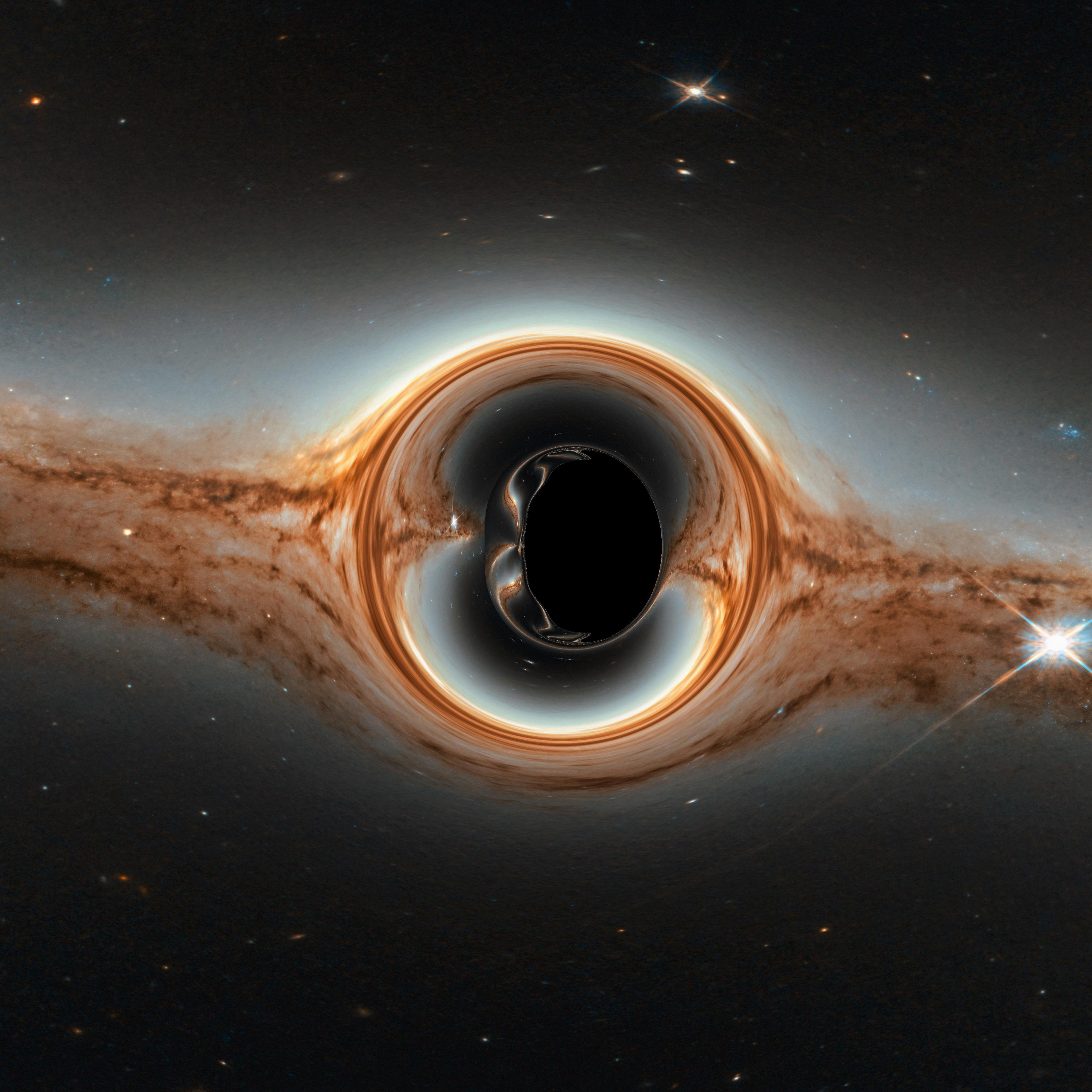}
	\includegraphics[width=0.46\textwidth]{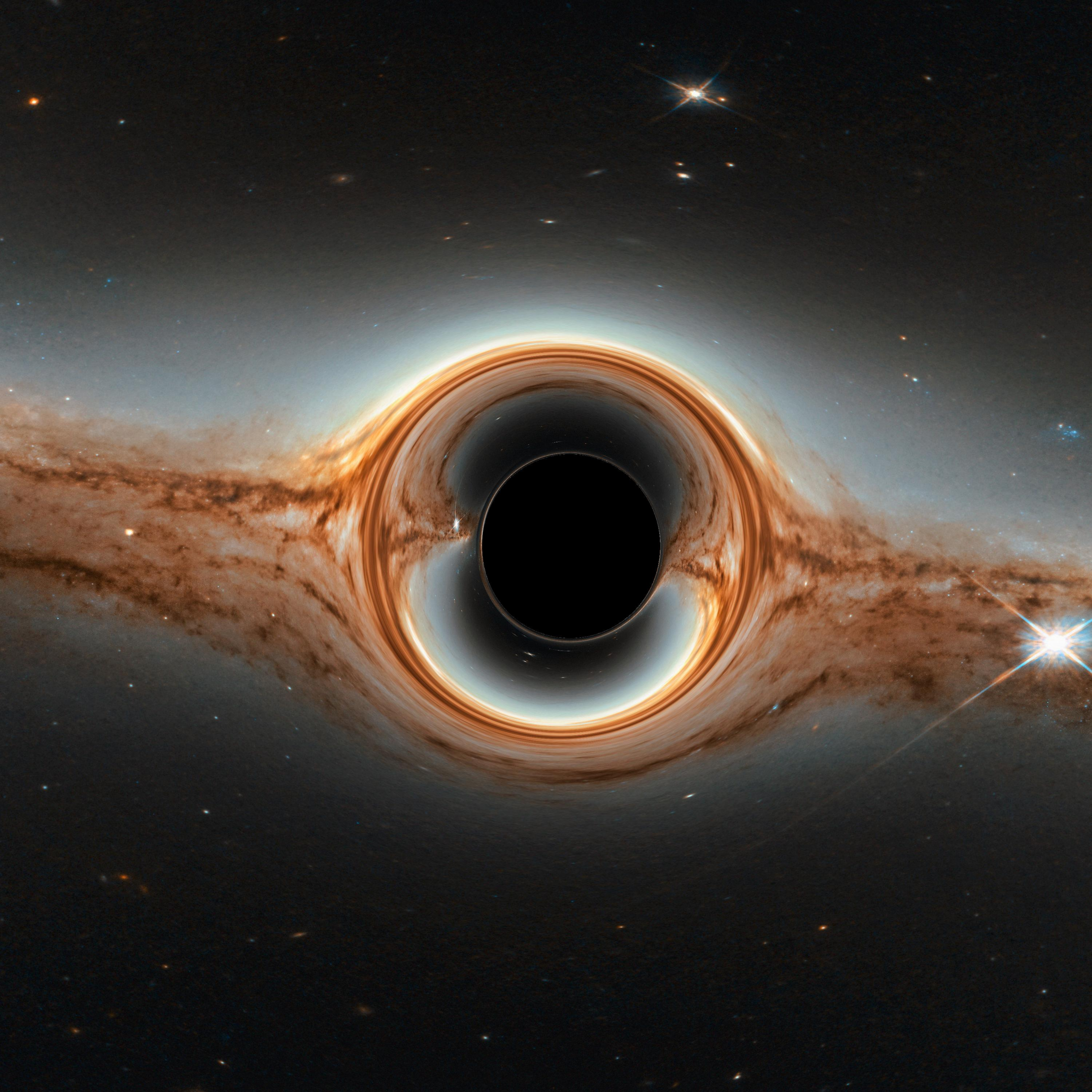}
	\caption{Images of HBH 3 (left) and a Kerr BH (right) with identical ADM mass and angular momentum for  $\theta_{\text{obs}}=90^\circ$. Background image adopted from \cite{hubble_image}.}
	\label{Fig: image}
\end{figure}

An advantage of this setup is that it enables detailed tracing of the origin of each light ray. Actually, the artificial background can be replaced with any alternative background image, allowing the generation of BH images in more astrophysical relevant contexts. An example is illustrated in Fig. \ref{Fig: image}, which presents the images of HBH 3 and a Kerr BH, both sharing identical ADM mass and angular momentum for an inclination angle of $\theta_{\text{obs}}=90^\circ$.

\section{Black hole shadow}
In this section, we analyse the shadows of HBHs 1-5 for an observer located at $\tilde{r}=30M_{\text{ADM}}$, where the circumferential radius is given by \cite{Cunha:2015yba}
\begin{equation}
	\tilde{r}=\frac{1}{2\pi}\int_{0}^{2\pi}\sqrt{g_{\varphi\varphi}}d\varphi.
\end{equation} 
Without stated otherwise, we choose $\theta_{\text{obs}}=90^\circ$.

We begin with HBHs 1 and 2, both of which have Kerr-like central horizons ($J_h/M^2_h<1$, see Table \ref{Table: Ref1}), as the scalar hair contributes relatively little to the total mass and angular momentum. For HBH 1, the scalar charge is $q=0.028$, with the scalar hair accounts for only $1.4\%$ of the total mass, while for HBH 2, the scalar charge is $q=0.304$, with the scalar hair constitutes $17.6\%$ of the total mass. 

\begin{table}
	\centering
	\caption{Physical quantities of the HBH solutions marked in Fig. \ref{Fig: param space}.}
	\begin{tabular}{@{}l*{7}{c}r@{}}
		\hline
		\hline
		& HBH  & $w$     & $M_{\text{ADM}}$ & $J_{\text{ADM}}$ & $q$     & $M_{\Psi}/M_{\text{ADM}}$ & $J_h/M^2_h$ & \\
		\hline
		& 1    & $0.985$ & $0.496$          & $0.243$          & $0.028$ & $1.4\%$                      & $0.991$ &\\
		& 2    & $0.980$ & $0.588$          & $0.334$          & $0.304$ & $17.6\%$                      & $0.988$ &\\
		& 3    & $0.801$ & $1.113$          & $0.845$          & $0.876$ & $84.7\%$                      & $3.618$ &\\
		& 4    & $0.747$ & $1.227$          & $0.985$          & $0.967$ & $91.5\%$                       & $3.009$ &\\
		& 5    & $0.700$ & $1.311$          & $1.153$          & $0.999$ & $97.7\%$                       & $1.037$ &\\
		\hline
		\hline
	\end{tabular}\label{Table: Ref1}
\end{table}

\begin{figure}
	\centering	
	\includegraphics[trim=80 5 80 5, clip, width=0.4\textwidth]{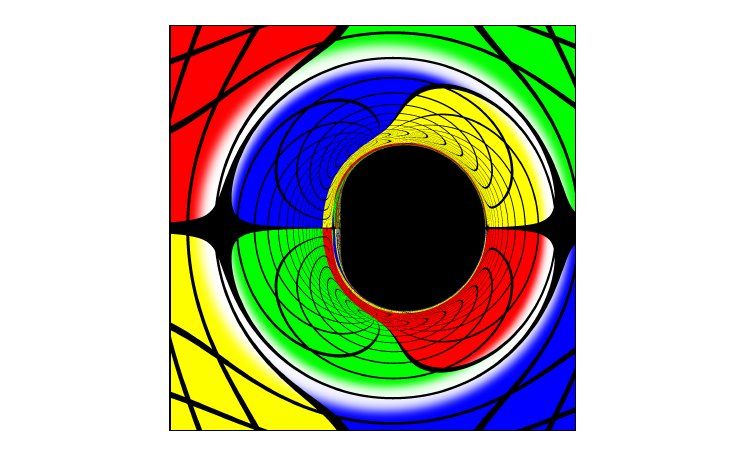}
	\includegraphics[trim=80 5 80 5, clip, width=0.4\textwidth]{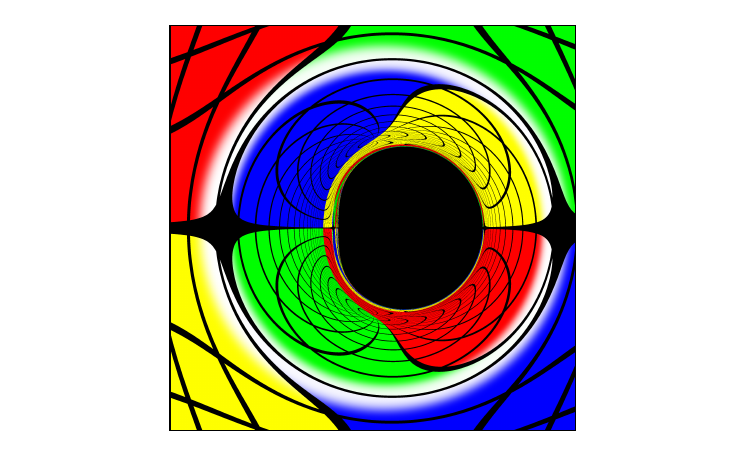}
	\includegraphics[trim=80 5 80 5, clip, width=0.4\textwidth]{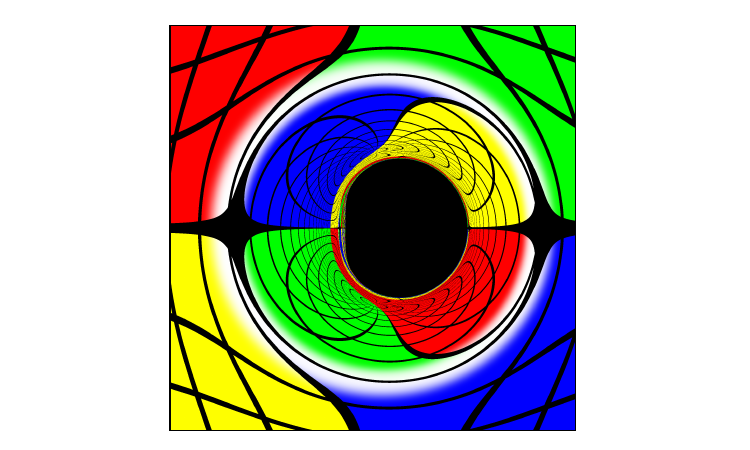}
	\includegraphics[trim=80 5 80 5, clip, width=0.4\textwidth]{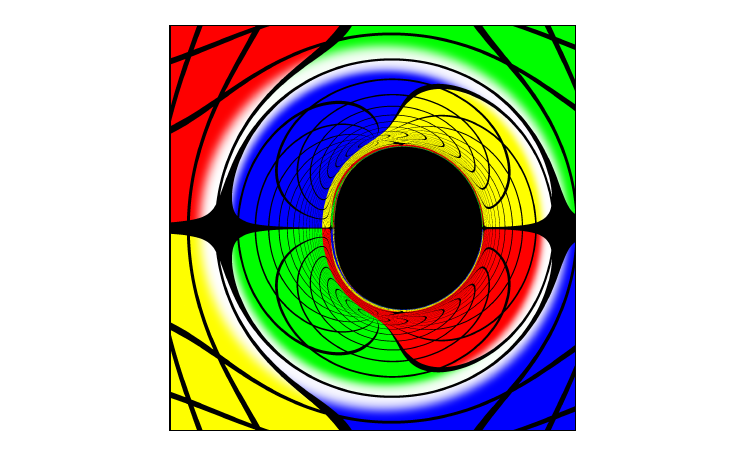}
	\caption{Top: images of HBH 1 (left) and Kerr BH (right). Bottom: images of HBH 2 (left) and its Kerr counterpart (right). Both Kerr BHs have identical ADM mass and angular momentum as their corresponding HBHs.}
	\label{Fig: Kerr}
\end{figure}

In Fig. \ref{Fig: Kerr}, we compare the shadows and lensing effects of HBHs 1 and 2 with those of Kerr BHs sharing the same ADM mass and angular momentum. The top row illustrates HBH 1 alongside its Kerr counterpart, while the bottom row presents HBH 2 and its corresponding Kerr BH. It can be observed that the images of HBHs 1 and 2 closely resemble those of their corresponding Kerr BHs. Both the shadows of HBHs and Kerr BHs shift to the right and exhibit a "D shape". However, subtle differences exist, particularly for HBH 2, whose shadow is noticeably smaller than that of its Kerr counterpart (see bottom row of Fig. \ref{Fig: Kerr}). 

To quantify the differences between the shadows of HBHs and Kerr BHs, following \cite{Cunha:2015yba,Johannsen:2013vgc}, we introduce six parameters $\{D_C,D_x,D_y,\bar{r},\sigma_r,\sigma_K \}$. Let the image parametrized by the Cartesian coordinates $(x,y)$, with the origin points at the (unlensed) white dot on the celestial sphere. The center of the shadow has $x_C=(x_\text{max}+x_\text{min})/2$, where $x_\text{max}$ and $x_\text{min}$ ($y_\text{max}$ and $y_\text{min}$) are the maximum and minimum abscissae (ordinates) of the shadow’s edge, respectively. The displacement of the shadow center from the image center is defined as $D_C\equiv|x_C|$. The width and height of the shadow are $D_x\equiv x_\text{max}-x_\text{min}$ and $D_y\equiv y_\text{max}-y_\text{min}$, respectively. The average radius of the shadow with respect to its center is given by
\begin{equation}
	\bar{r}=\frac{1}{2\pi}\int_{0}^{2\pi} r(\vartheta)d\vartheta,
\end{equation}
where $r(\vartheta)\equiv\sqrt{(x-x_C)^2+y^2}$ and $\tan\vartheta\equiv y/(x-x_C)$. The deviation from sphericity is given by
\begin{equation}
	\sigma_r\equiv\sqrt{\frac{1}{2\pi}\int_{0}^{2\pi}\left[\frac{r(\vartheta)-\bar{r}}{\bar{r}}\right]^2d\vartheta}.
\end{equation}
Finally, the relative deviation from a comparable Kerr BH is given by
\begin{equation}
	\sigma_K\equiv\sqrt{\frac{1}{2\pi}\int_{0}^{2\pi}\left[\frac{r(\vartheta)-r_{\text{Kerr}}(\vartheta)}{r_{\text{Kerr}}(\vartheta)}\right]^2d\vartheta}.
\end{equation}

\begin{figure}
	\includegraphics[trim=40 190 40 190, clip, width=0.46\textwidth]{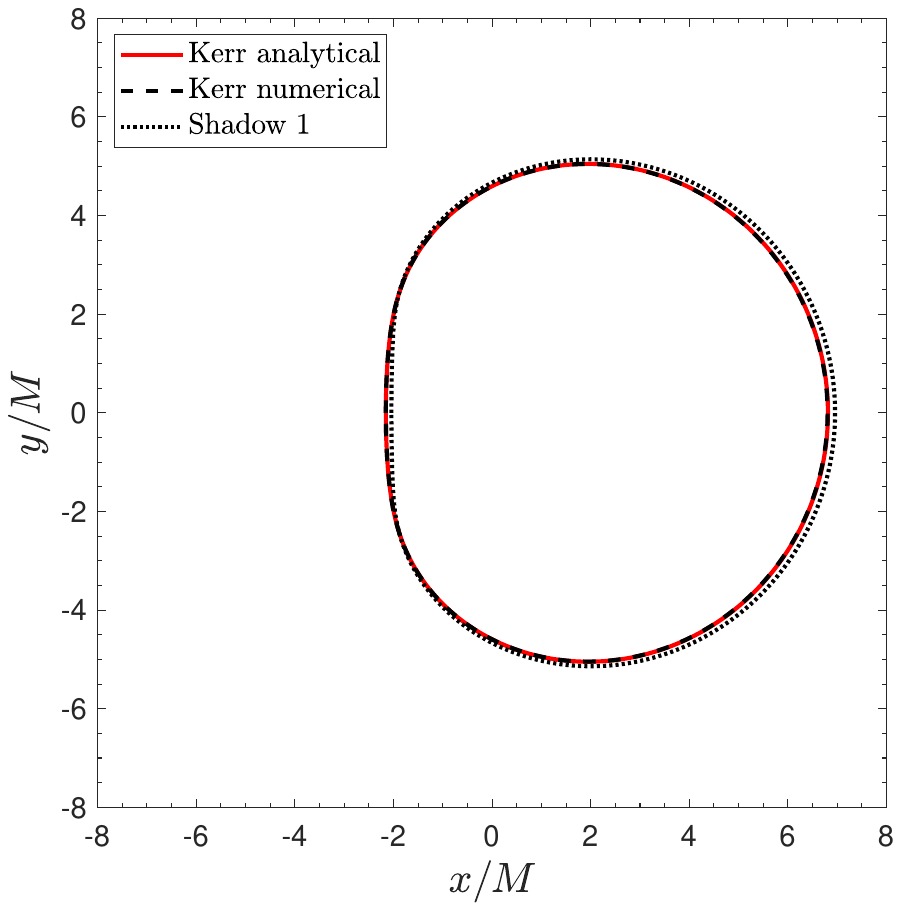}
	\includegraphics[trim=40 190 40 190, clip, width=0.46\textwidth]{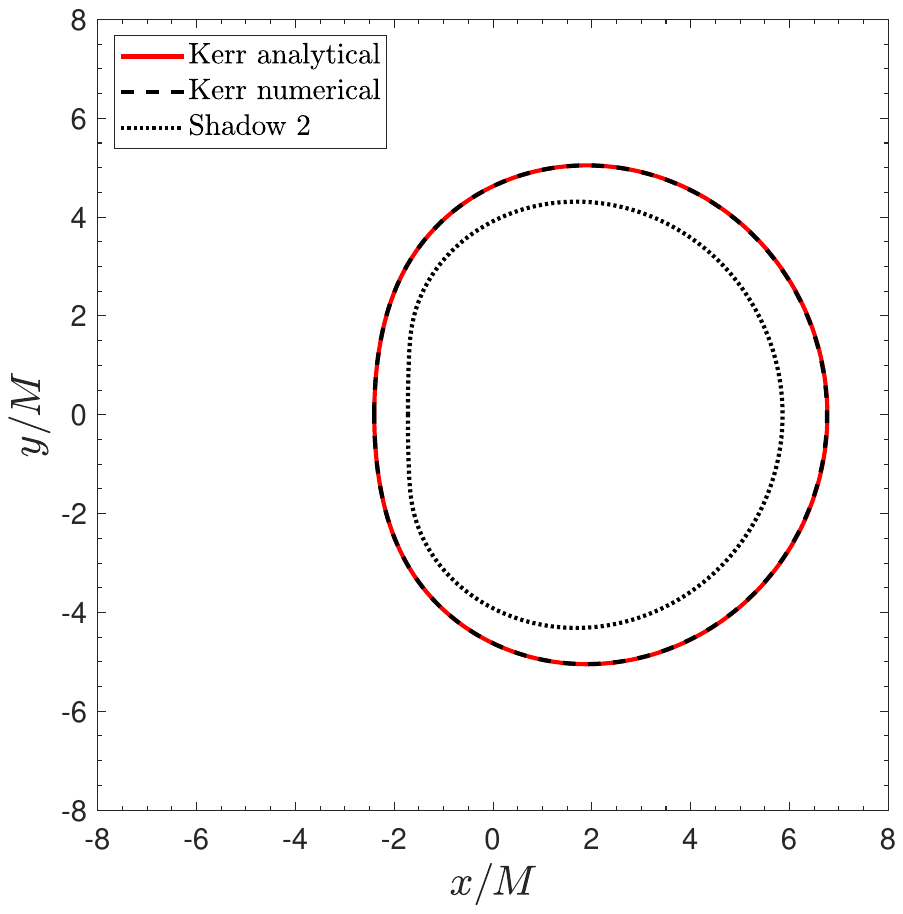}
	\caption{Shadows of HBH 1 (left) and HBH 2 (right) compared with their Kerr counterparts. Numerical results for the Kerr BHs are generated using the same ray tracing code as for the HBHs, while 
		analytical solutions are from \cite{Cunha:2016bpi}.}
	\label{Fig: shadow compare}
\end{figure}

We compare the shadows of HBHs 1, 2, and their Kerr counterparts in Fig. \ref{Fig: shadow compare}, and list the corresponding shadow parameters in Table \ref{Tab: Parameters}. For the Kerr BHs, we present both numerical results (generated using the same ray tracing code as for the HBHs) and the analytical solutions from \cite{Cunha:2016bpi}. From Fig. \ref{Fig: shadow compare} and Table II, it is evident that the shadows of HBHs 1 and 2 closely resemble their Kerr counterparts, though with notable quantitative differences. Specifically, HBH 1 exhibits a deviation of approximately $1.64\%$ from its comparable Kerr shadow, while HBH 2 shows a more pronounced difference of $15.44\%$. In addition, the numerical results for the Kerr shadows excellently agree with the analytical solutions, with discrepancies below $0.05\%$, demonstrating the high precision and validity of our code.

\begin{table*}
	\centering
	\caption{Parameters of shadows 1 and 2 with their corresponding Kerr shadows of matching ADM masses and angular momenta, including both numerical results (generated using the same ray tracing code as for the HBHs) and analytical results for an observer located at $\tilde{r}=30M_{\text{ADM}}$. The numerical and analytical Kerr shadows show excellent agreement.}
	\setlength{\tabcolsep}{10pt}
	\begin{tabular}{@{}c c c c c c c c c@{}}
		\hline
		\hline
		&          & $D_C$   & $D_x$   & $D_y$   & $\bar{r}$ & $\sigma_r(\%)$ & $\sigma_K(\%)$ & \\
		\hline
		& Shadow 1 & $2.457$ & $8.997$ & $10.28$ & $4.882$   & $5.660$ 	      & $1.647$        & \\
		& Kerr num & $2.325$ & $8.958$ & $10.09$ & $4.809$   & $4.938$ 		  & $0.054$ 	   & \\
		& Kerr ana & $2.329$ & $8.965$ & $10.09$ & $4.812$   & $4.925$ 		  & $0.000$	   	   & \\
		\hline
		& Shadow 2 & $2.068$ & $7.582$ & $8.626$ & $4.104$   & $5.552$ 		  & $15.44$        & \\
		& Kerr num & $2.177$ & $9.163$ & $10.09$ & $4.844$   & $3.870$ 		  & $0.049$ 	   & \\
		& Kerr ana & $2.181$ & $9.169$ & $10.09$ & $4.846$   & $3.858$ 		  & $0.000$	   	   & \\
		\hline
		\hline
	\end{tabular}\label{Tab: Parameters}
\end{table*}

\begin{figure}
	\centering	
	\includegraphics[trim=80 10 80 10, clip, width=0.44\textwidth]{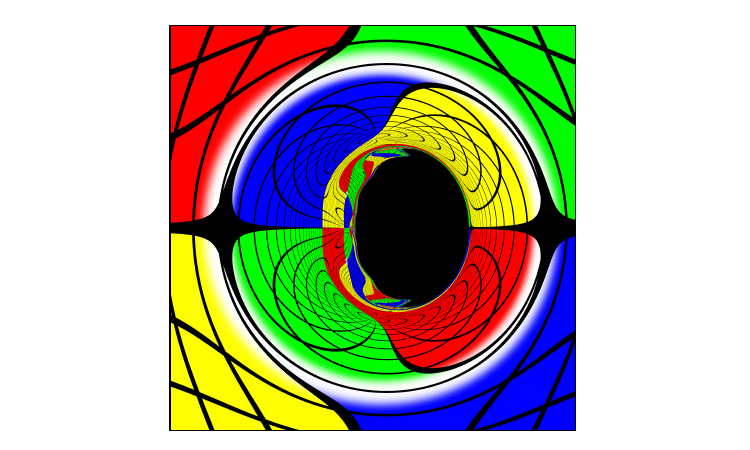}
	\includegraphics[trim=80 10 80 10, clip, width=0.44\textwidth]{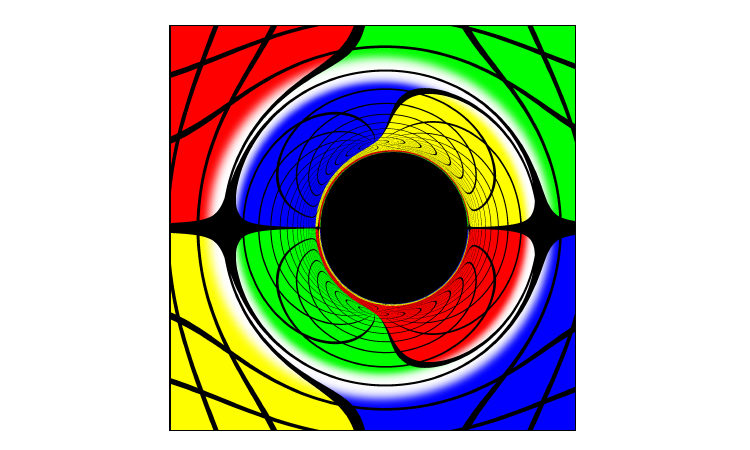}
	\includegraphics[trim=80 10 80 10, clip, width=0.215\textwidth]{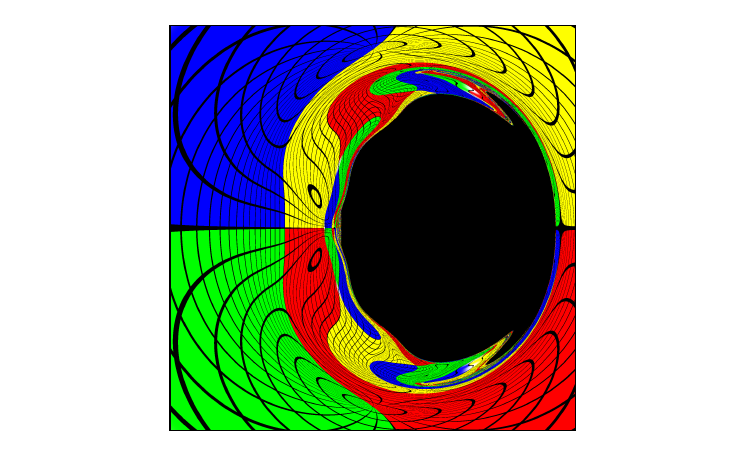}
	\includegraphics[trim=80 10 80 10, clip, width=0.215\textwidth]{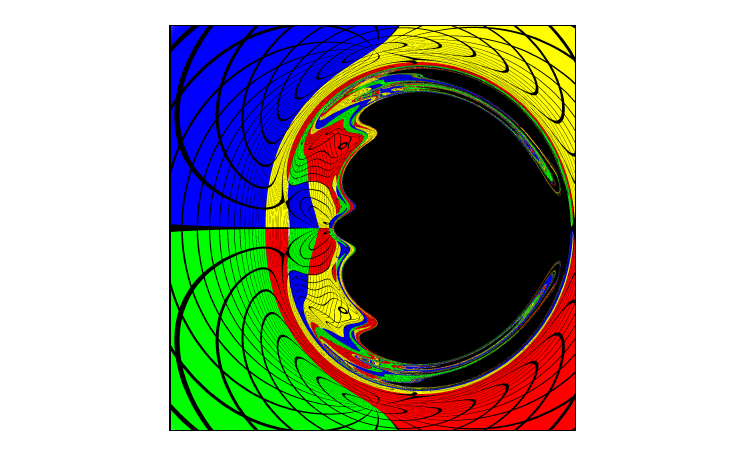}
	\includegraphics[trim=80 10 80 10, clip, width=0.215\textwidth]{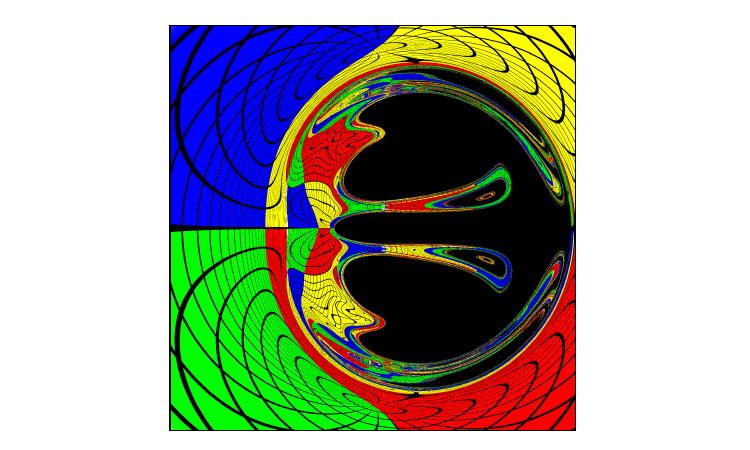}
	\includegraphics[trim=80 10 80 10, clip, width=0.215\textwidth]{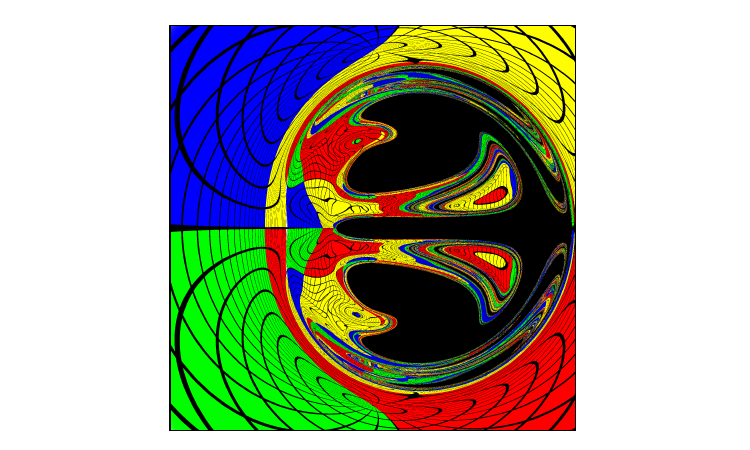}
	\includegraphics[trim=80 10 80 10, clip, width=0.44\textwidth]{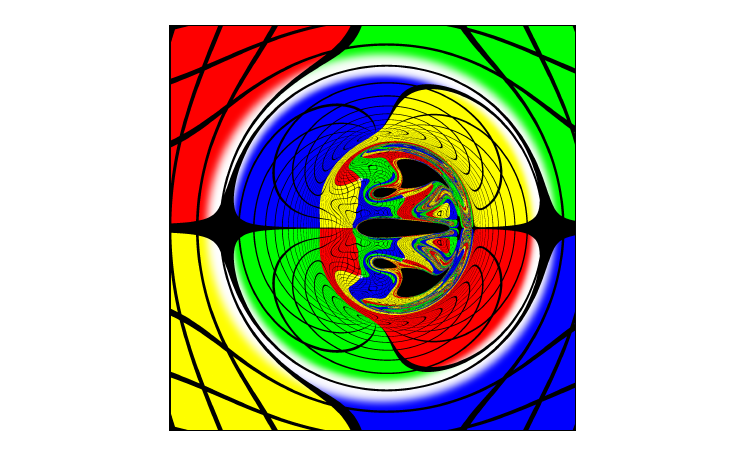}
	\includegraphics[trim=80 10 80 10, clip, width=0.44\textwidth]{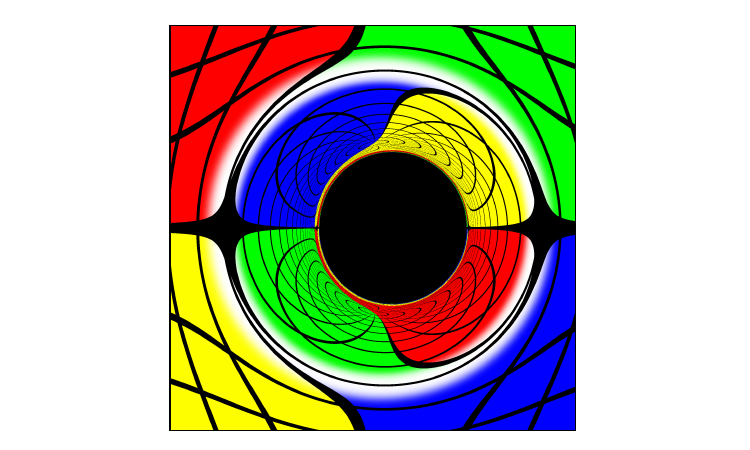}
	\caption{Shadows for HBH 3 (top left) and HBH 4 (bottom left) compared with their Kerr counterparts in the corresponding right panels. The middle row shows four transition examples.}
	\label{Fig: BH lensing1}
\end{figure}

Let us now turn to the cases where the scalar hair induces significant deviations from the Kerr solution and distinct shadow features appear. In Fig. \ref{Fig: BH lensing1}, we present the images for HBH 3 (top left) and HBH 4 (bottom left), along with their Kerr counterparts in the corresponding right panels. In addition, the middle row illustrates four transition examples between the two HBHs. We see that both images 3 and 4 exhibit significant deviations from the Kerr cases: while image 3, like that of Kerr, consists of a single primary shadow, it is notably flatter and exhibits sharp peaks in some regions, whereas image 4 shows at least five disconnected sub-shadows despite having only a single event horizon, with even smaller eyebrows-like shadows observed in the image\textemdash a phenomenon previously seen in binary BH systems and even-parity HBHs \cite{Cunha:2015yba,Bohn:2014xxa}. The results in Fig. \ref{Fig: BH lensing1} thus provide a novel example demonstrating that a single event horizon can generate multiple shadows. 

More interestingly, in the second row of Fig. \ref{Fig: BH lensing1}, the first two images exhibit shadows that bear a striking resemblance to human side face. This type of shadow has been previously reported in the context of a slowly rotating BH within a modified theory of gravity, where an axion field is coupled to the Chern-Simons term \cite{Kuang:2024ugn}. This observation suggests that odd-parity HBHs can effectively mimic the shadow features of BHs in more complex modified gravity theories. Remarkably, despite the simplicity of the theoretical framework under consideration\textemdash a minimally coupled massive scalar field to gravity\textemdash the resulting odd-parity solutions are capable of reproducing intricate shadow patterns typically associated with highly non-trivial gravitational modifications.

\begin{figure}
	\centering	
	\includegraphics[trim=80 10 80 10, clip, width=0.3\textwidth]{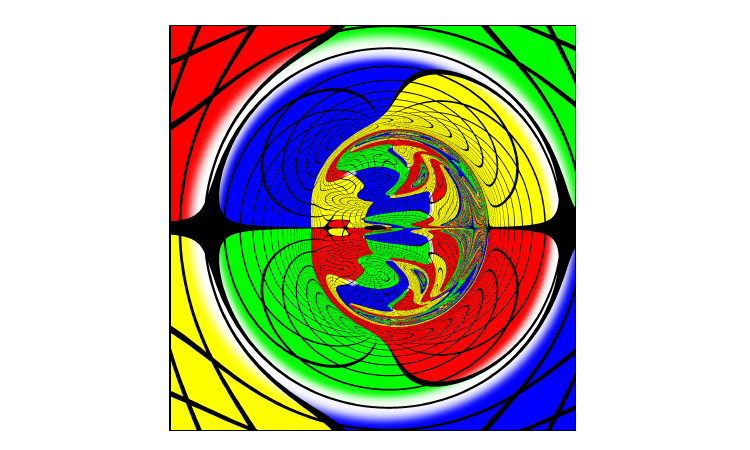}
	\includegraphics[trim=80 10 80 10, clip, width=0.3\textwidth]{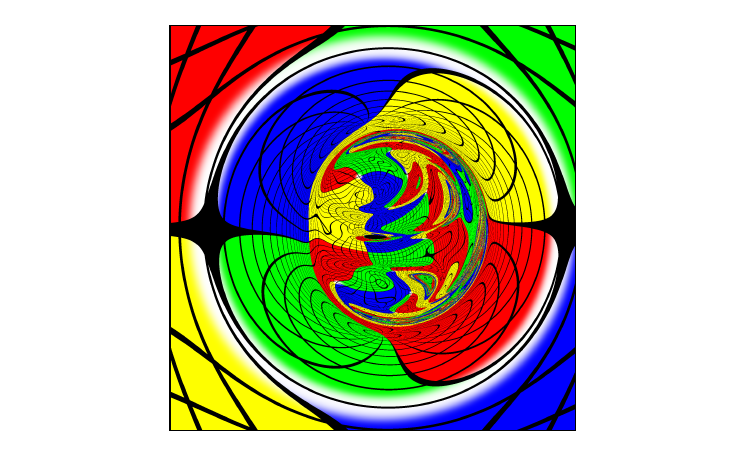}
	\includegraphics[trim=80 10 80 10, clip, width=0.3\textwidth]{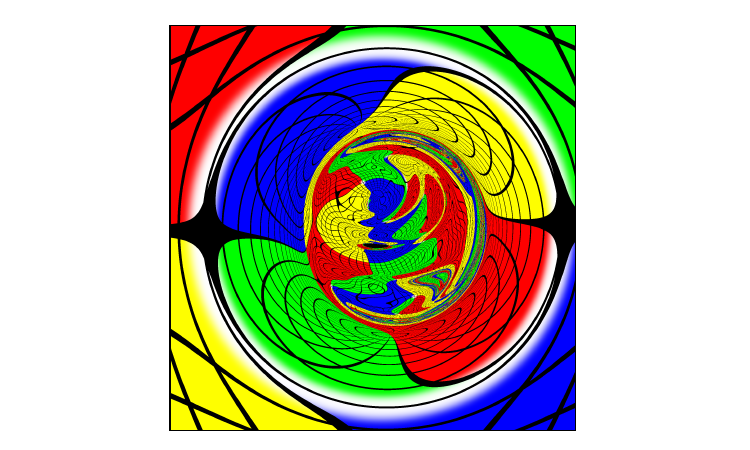}
	\includegraphics[trim=80 10 80 10, clip, width=0.3\textwidth]{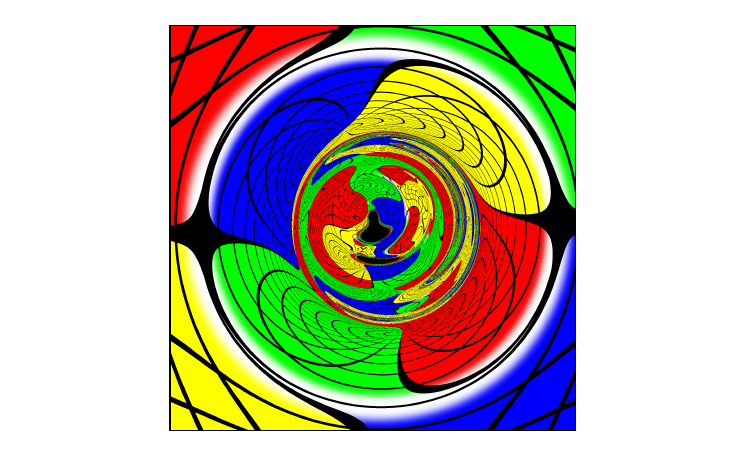}
	\includegraphics[trim=80 10 80 10, clip, width=0.3\textwidth]{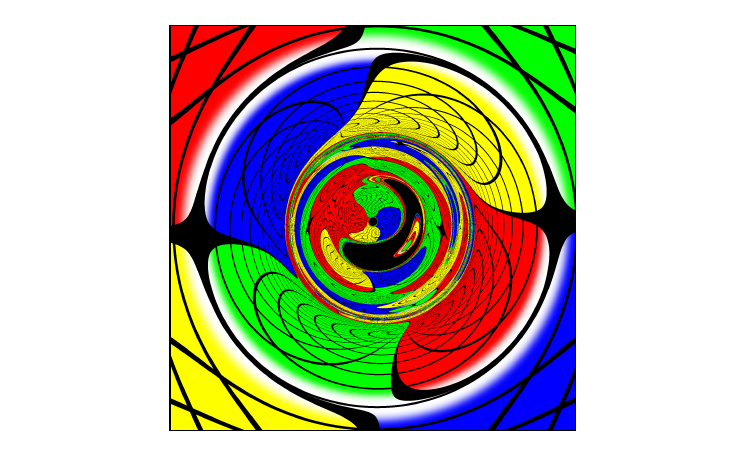}
	\includegraphics[trim=80 10 80 10, clip, width=0.3\textwidth]{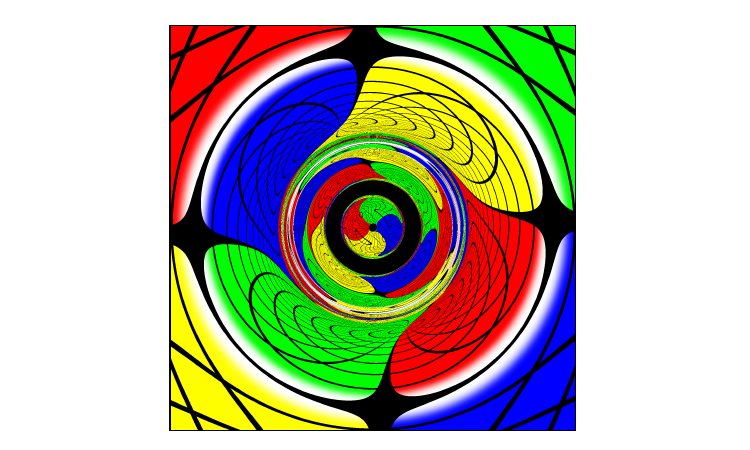}
	\includegraphics[trim=80 10 80 10, clip, width=0.3\textwidth]{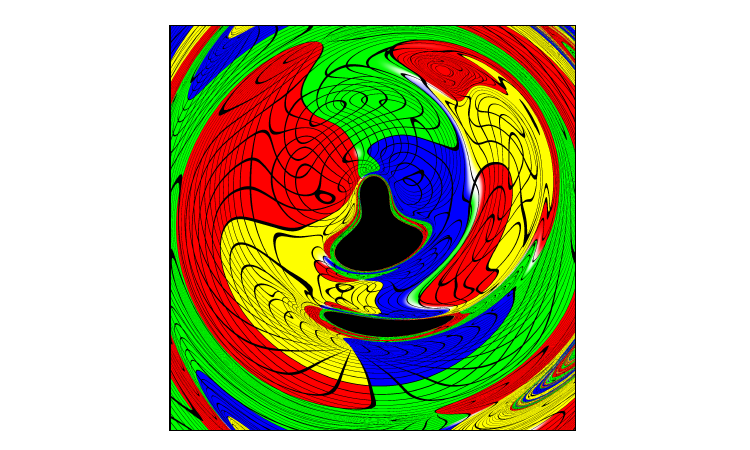}
	\includegraphics[trim=80 10 80 10, clip, width=0.3\textwidth]{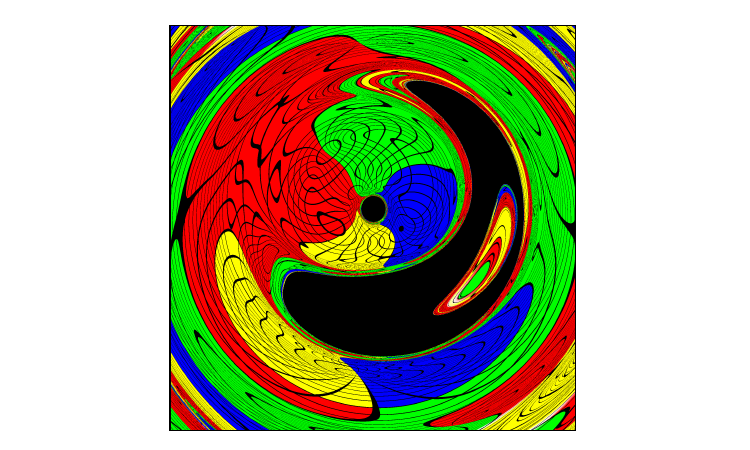}
	\includegraphics[trim=80 10 80 10, clip, width=0.3\textwidth]{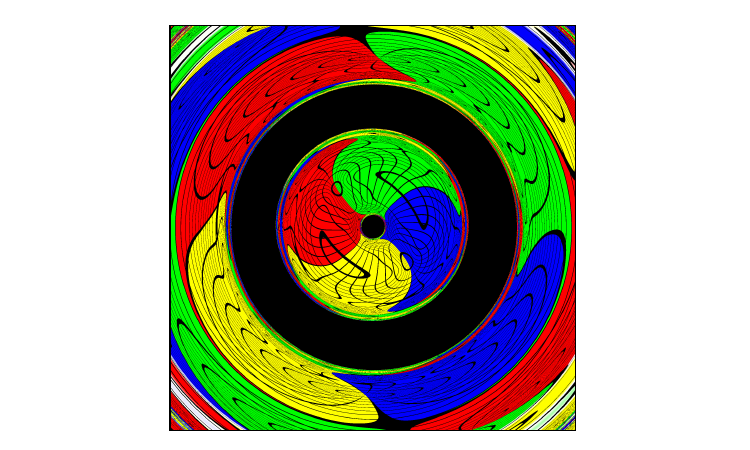}
	\caption{Images of HBH 5 at different observer inclination angles. From left to right, $\theta_{\text{obs}}=90^\circ, 75^\circ, 60^\circ$ (top row), $\theta_{\text{obs}}=30^\circ, 15^\circ, 
		1^\circ$ (middle row), and zoomed-in views of the middle row images (bottom row).}
	\label{Fig: BH lensing2}
\end{figure}

Finally, we focus on HBH 5, where the scalar hair dominates the physical effects of the background spacetime, as it contributes about $97.7\%$ to the total ADM mass and the scalar charge reaches $q=0.999$\textemdash the highest among all cases studied, the spacetime is primarily shaped by the scalar field distribution, with the central BH playing a negligible role. Figure \ref{Fig: BH lensing2} shows that the images are nearly identical to those of the boson stars in \cite{Huang:2024gtu}. The shadow generated by the event horizon at the center is extremely small, particularly at $\theta_{\text{obs}}=90^\circ$, where it appears as a tiny, nearly imperceptible droplet shape. Notably, as the observer’s inclination angle decreases, the size of the shadow gradually increases, and its shape undergoes significant changes. Starting from  $\theta_{\text{obs}}=30^\circ$, the shadow splits into two distinct components. By $\theta_{\text{obs}}=15^\circ$, one component becomes circular disk, while the other transforms into a ring-like structure, with the disk located outside the ring. Finally, at $\theta_{\text{obs}}=1^\circ$, the ring-like shadow encloses the smaller disk at its center.

\begin{figure}
	\centering	
	\includegraphics[trim=80 5 80 5, clip, width=0.325\textwidth]{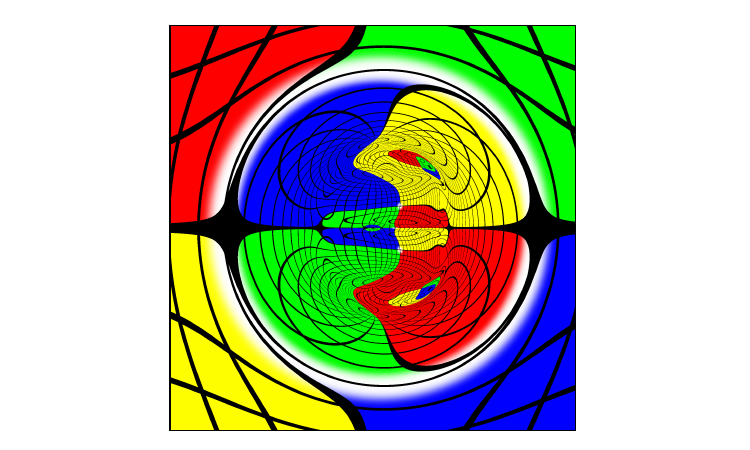}
	\includegraphics[trim=80 5 80 5, clip, width=0.325\textwidth]{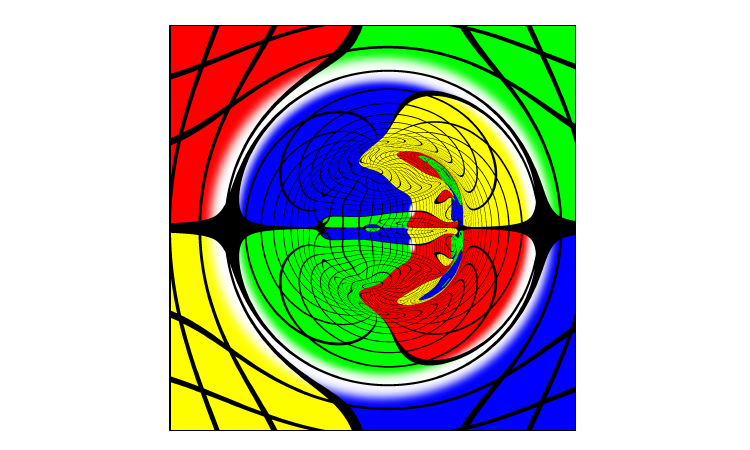}
	\includegraphics[trim=80 5 80 5, clip, width=0.325\textwidth]{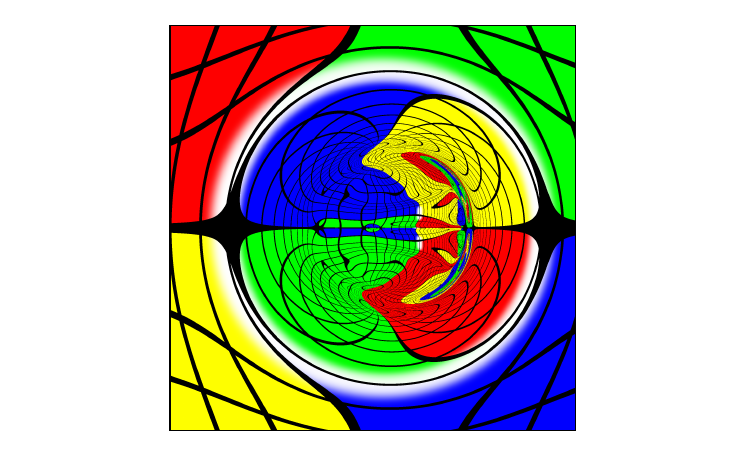}
	\caption{Hairy black hole shadows for $m=1$ (left), $2$ (middle) and $3$ (right), with the frequency $w=0.745$ and the horizon radius $r_h=0.005$.}
	\label{Fig: shadow3}
\end{figure}

  In the above analysis, we selected $m=1$ as the representative case for simplicity, as it captures the fundamental features of the rotating hairy black hole. In Fig. \ref{Fig: shadow3}, we further study the hairy black hole shadows for different values of $m$. While varying $m$ quantitatively alters the shadow's morphology by increasing the number of replicated quadrants (as shown in Fig. \ref{Fig: shadow3}), the qualitative behavior remains fundamentally unchanged. Specifically, all cases consistently exhibit the characteristic parity-odd solution properties: the number of copies is exactly twice that of parity-even solutions, as demonstrated in \cite{Huang:2024gtu}. 

\begin{figure}
	\centering	
	\includegraphics[width=0.6\textwidth]{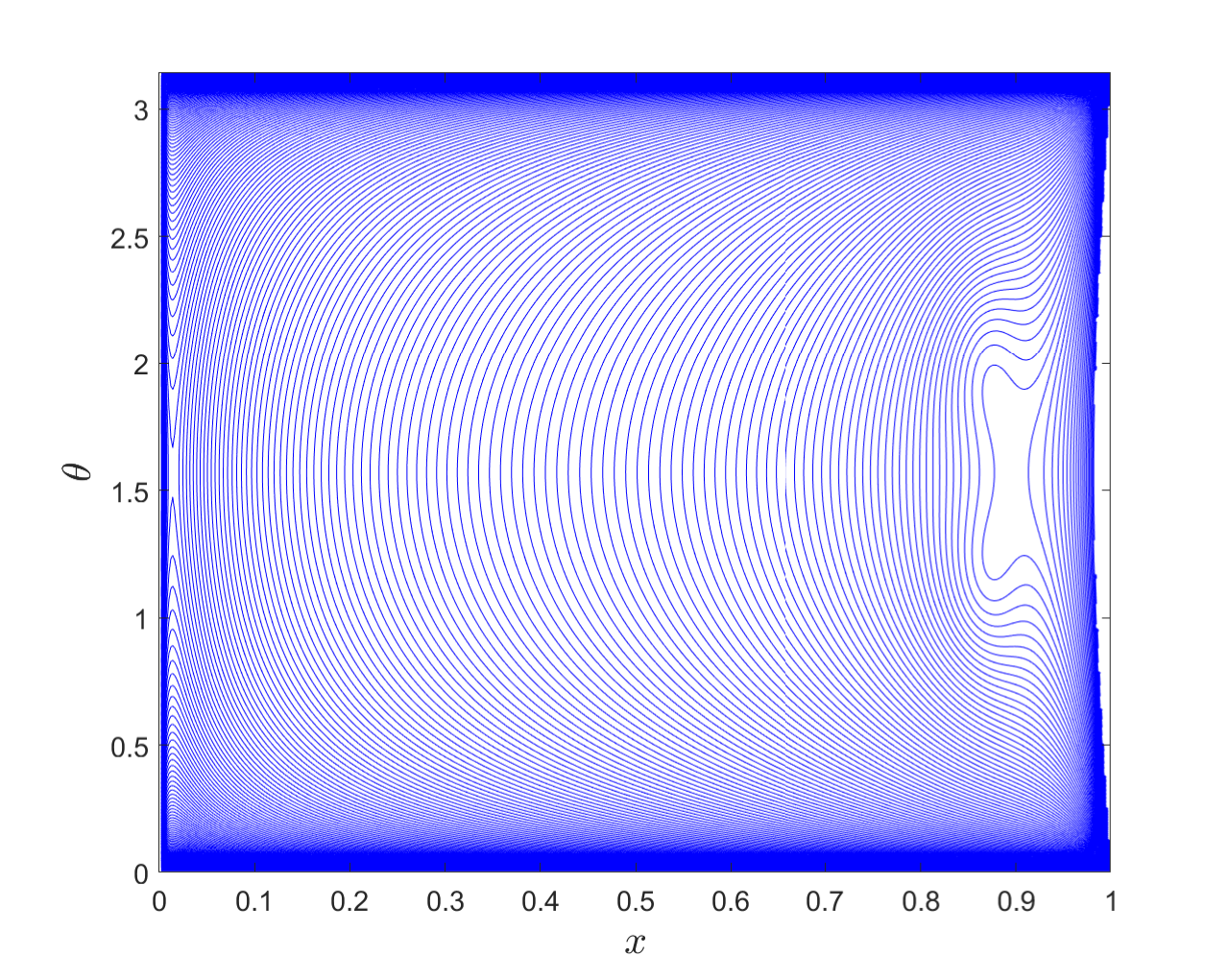}
	\caption{Contour plot of $h_+$ for the hairy black hole with $m=3$ given in Fig. \ref{Fig: shadow3}.}
	\label{Fig: potential2}
\end{figure}

Focus on the $m=3$ case, the image clearly exhibits chaotic patterns in mirror-symmetric positions relative to the equatorial plane, demonstrating parity-odd solution properties. This symmetric distribution of chaotic regions originates from the stable light rings characteristic of parity-odd solutions, which naturally occur in pairs symmetric about the equatorial plane. In Fig. \ref{Fig: potential2}, we present the potential $h_+$ of the $m=3$ solution given in Fig. \ref{Fig: shadow3}. The stable light rings are associated with the local minima of the effective potential $h_+$ \cite{Cunha:2016bjh,Huang:2024gtu}. As clearly shown, the potential exhibits two minima located symmetrically across the equatorial plane.

\section {Conclusion}
We investigate the shadows of HBHs and compare them with those of Kerr BHs. Our findings reveal that the shadows of HBHs are distinguishable from those of Kerr BHs. When the scalar charge $q$ is very small and the mass contribution of the scalar hair is minimal, the differences between HBH and Kerr shadows are subtle but quantifiable. For instance, in the case of HBH 1, where $q=0.028$ and the scalar hair contributes only $1.4\%$ to the total ADM mass, the deviation from Kerr is $1.64\%$. This solution lies within the error bars of the EHT observations. We thus obtain that  the scalar mass to about $1.02\times10^{-20}$eV from the image of M87*.

When the scalar charge is very large and the mass contribution of the scalar hair becomes substantial, the HBH shadows exhibit striking deviations from Kerr shadows, displaying novel and intricate shapes. One notable feature is the potential fragmentation of the HBH shadow into multiple disconnected components, each with distinct shapes such as teardrop or eyebrow-like forms. Another intriguing observation is the emergence of "human-face-like" shadows, previously found in Chern-Simons gravity \cite{Kuang:2024ugn}. The results in the present work show that parity-odd HBHs effectively mimic the shadow characteristics of BHs in more complex modified gravity theories.

Our results provide new templates for BH shadows, particularly highlighting the influence of parity-odd matter distributions on BH shadow morphology. These findings may have profound implications for future observations, offering new avenues to explore the nature of BHs and the underlying theories of gravity.

\acknowledgments
 
We are grateful to the anonymous reviewer, whose valuable comments and suggestions significantly enhanced the quality of this paper.  This work was supported by the National Natural Science Foundation of China (Grants Nos. 12275106 and 12235019), the Shandong Provincial Natural Science Foundation (Grant No. ZR2024QA032), Youth Innovation Group Plan of Shandong Province (Grant No. 2023KJ107), and the Innovation Grogram of Shanghai Normal University (Grant No. KF202472).


\end{document}